\documentstyle[multicol,aps,prb,tabularx,epsf]{revtex}

\begin{document}
\title{Effect of quantum interference in the \ nonlinear conductance of
microconstrictions}
\author{A. Namiranian$^{ 1 }$, Yu.A. Kolesnichenko$^{ 1,2 }$%
, A.N. Omelyanchouk$^{ 2 }$}

\address{$1$ Institute for Advanced Studies in Basic Sciences, Gava Zang,
Zanjan   45195-159, Iran \\
$2$ B.Verkin Institute for Low Temperature Physics and Engineering,\\ National
Academy of Sciences of Ukraine,\\ 47 Lenin Ave., 310164 Kharkov, Ukraine}

\date{\today}
\maketitle
\begin{abstract}
The influence of the interference of electron waves, which were scattered by
single impurities, on nonlinear quantum conductance of metallic microconstrictions
(as was recently investigated experimentally \quad \cite{Rutenbeek,Agrait} )
is studied theoretically.\ The dependence of the interference
pattern in the conductance $G\left( V\right) $ on the contact diameter and
the spatial distribution of impurities is analyzed. It is shown that the
amplitude of conductance oscillation is strongly depended on the position of
impurities inside the constriction.
\end{abstract}

\begin{multicols}{2}

\section{Introduction.}

\bigskip The Scanning Tunneling Microscopy (STM) and the Mechanically
Controllable Break-junction (MCB) techniques offer an opportunity to study
the conductance of metallic contacts consisting of only a few atoms (quantum
contacts). The electrical conductance of such contacts, at small bias
voltage is proportional to the number of propagating electron modes, $N$,
each one contributing an amount of $G_{0}=2e^{2}/h \;  \;$\cite{SSP}. With
increasing the diameter of the contact the energies of modes continuously
decreases, but the number of modes increases whenever a new mode fits into
the constriction cross section. This number, $N,$ is limited by the
requirement that the kinetic energy for the transverse motion is smaller
than the Fermi energy $\varepsilon _{F}$. When a new mode is occupied, a new
quantum channel is opened. The conductance then undergoes a jump of $G_{0}$.
Such quantization of conductance has been observed in both two and three
dimensional contacts with diameters comparable to Fermi wave-length $\lambda
_{F}=h/p_{F}$ ($p_{F}$ is the Fermi momentum) \cite{2Dfirst,2Dfirst-too,STM1,STM2,BJ1}
. Jumps in conductance
are also expected to occur, at the constant contact diameter while bias
voltage is varied. If the bias $eV$ is larger than the distances between the
energy levels of quantum modes, it is possible to change the number of
opened modes by changing the voltage $V.$ At a certain threshold voltage a
channel is opened or closed for one direction of the electron wave vector
along the constriction and consequently conductance suffers a $G_{0}/2$
stepwise change \cite{Zagoskin}.

Quantum interference effects have been studied in different
mesoscopic systems \cite{Imry}. In ballistic metallic
microconstrictions it manifests itself as fluctuations in
conductance when a magnetic field \cite{Hfluct} or an electrical
voltage is applied \cite{Vfluct}. Now experimental efforts have
been done using MCB techniques, in order to measure conductance as
a function of voltage in atomic-size point contacts
\cite{Rutenbeek}. A prominent feature of these measurements, is
the existence of small random voltage-dependent fluctuations in
conductance, far from steps. The measurements \cite{Rutenbeek}
clearly indicate suppression of the fluctuations for conductance
values near the integer multiples of the conductance quantum.
Similar results have been reported by using a STM to show the
strong voltage dependence of conductance of one-atom contacts at
different temperatures \cite{Agrait}. It is generally believed
that the observed oscillations in conductance are due to the
quantum interference effects \cite{Agrait}. Ludoph and co-authors
\cite{Rutenbeek}, propose the following interpretation: The electron
wave transmitted through the contact is backscattered to the
contact by an impurity and then partially reflected at the
contact. These waves interfere and change
the total conductivity. The energy and thus the wave
number of an injected electron into the channel, depends on the
voltage. Consequently the interference pattern in conductance
oscillates as electron wave number varies with the voltage.

Although the theory developed by Ludoph et al. \cite{Rutenbeek}
can explain the general feature of fluctuations but here we try to
examine a different mechanism. Impurities (or defects) are assumed
to be located inside the constriction, and the interference is
effectively between waves scattered
from the impurities. The existence of a few defects or impurities
inside the constriction is rather natural considering the way the
contact is formed. Using the model of a long microconstriction we
can find the conductivity analytically. We discuss the
theory of nonlinear electron transport through a mesoscopic
microconstriction with a few impurities. We show that the
nonlinear dependence of the quantum conductance on the voltage is
obtained from this model. The form of this dependence is affected
not only by the distances between impurities, but also by their
positions inside the constriction.

In Sec. II the model Hamiltonian is discussed and is used to
obtain a general expression for the nonlinear conductance. In Sec.
III a $\delta $-function potential is assumed for the interaction
of electrons with impurities and a simplified equation for the
conductance is obtained. Within the framework of perturbation
theory, a general analytical equation is also derived for
conductance of the system, for arbitrary number of quantum modes
and arbitrary number of impurities located in arbitrary positions.
These analytical results are illustrated by numerical calculations
for the contact in the form of a long cylindrical contact.A brief
discussion of result is given in Sec. IV.

\section{General equation for the nonlinear conductance of the long
quantum microconstriction.}

\bigskip Let us consider a long narrow constriction, which connects two
bulk metals, assuming $eV\ll \varepsilon _{F}$. The geometry is
shown in Fig. 1. We assume that the contact shape is smooth on the
scale of the wavelength $\lambda _{F}.$ This condition assures
that different transverse modes pass through the ballistic contact
independently (adiabatic approximation \cite{Glazman}). We also
assume that the contact length is much larger than its diameter
and we can neglect the constriction end effects. Under these
approximations, the electrical field inside the contact far from
the ends is negligible and the energy $\varepsilon $ of ballistic
electrons depends only on the sign of velocity along the contact
axis \cite{KOSh,KulYan}. The Hamiltonian $H$ of the electrons
contains the following terms:
\begin{equation}
H=H_{0}+H_{1}+H_{int},
\end{equation}
where
\begin{equation}
H_{0}=%
\mathrel{\mathop{\sum}\limits_{n}}%
\varepsilon_{\alpha}c_{\alpha}^{\dagger}c_{\alpha}
\end{equation}
\noindent is Hamiltonian of free electrons, and
\begin{equation}
H_{1}=\frac{eV}{2}%
\mathrel{\mathop{\sum }\limits_{\alpha }}%
signv_{z}c_{\alpha }^{\dagger }c_{\alpha }
\end{equation}
\noindent describes the influence of applied bias voltage.
$H_{int}$ denotes interaction of electrons with impurities, and
depends on
the positions of impurities ${\bf r}_{i}$ in the constriction; the operator $%
c_{\alpha }^{+}\left( c_{\alpha }\right) $ creates (annihilates) a
conduction electron with the wave function $\varphi _{\alpha },$ and energy $%
\varepsilon _{\alpha }.$ The electron wave functions and eigenvalues are

\begin{eqnarray}
\varphi _{\alpha }({\bf r})& =&\psi _{\beta }({\bf R})\exp \left( \frac{i}{%
\hbar }p_{z}z\right) ;  \\
&\varepsilon _{\alpha }& =\varepsilon _{\beta }+\frac{p_{z}^{2}}{2m_{e}};
\end{eqnarray}

where $\alpha =\left( \beta ,p_{z}\right) ,$ $\beta $ is the set
of two transverse quantum numbers; $p_{z}$ is the momentum of an
electron along the contact axis. ${\bf r=}\left( {\bf R,}z\right)
,$ ${\bf R}$ is a coordinate in the plain, perpendicular to the $z$ axis $%
m_{e}$ is the electron mass.

In zero approximation in $H_{int}$ the current $J_{0}$ through the
contact area $S_{c}$ is

\begin{equation}
J_{0}=eS_{c}Tr\left( v_{z}\rho _{0}\right) ;
\end{equation}
where
\begin{equation}
\rho _{0}=f_{F}\left( H_{0}+H_{1}\right) ;
\end{equation}
$v_{z}=p_{z}/m$ is the electron velocity; $f_{F}$ is the Fermi
function. Using the Eqs.(6),(7), and wavefunctions (4), we find
the equation for the ballistic conductance:
\begin{equation}
G_{1}=\frac{1}{2}G_{0}\sum_{\beta }\left[ f_{F}\left( \varepsilon _{\beta }+%
\frac{eV}{2}\right) +f_{F}\left( \varepsilon _{\beta }-\frac{eV}{2}\right) %
\right] \quad
\end{equation}
At zero temperature and $V\rightarrow 0$, this formula describes the well known $%
G_{0}$ steps of quantum conductance and in the quasiclassical case
it turns into the Sharvin conductance \cite{Sharvin,KOSh}.

In order to investigate the influence of single impurities on the
nonlinear quantum conductance of the point contact, we use the
method, which was developed Kulik and others \cite{Kulik,KOT}. The
change in the electrical current $\Delta J$ is related to the rate
of energy dissipation by the relation:

\begin{equation}
\Delta JV=\frac{dE}{dt}=\frac{d\left\langle H_{1}\right\rangle }{dt};
\end{equation}

\bigskip Differential of $\left\langle H_{1}\right\rangle $ with respect
to time $t$ is we obtained from Heisenberg equation. The change
$\Delta J$ of the current due to interactions of electrons with
impurities; would then be

\begin{equation}
\Delta JV=\frac{1}{i\hbar }\left\langle \left[ H_{1}\left( t\right)
,H_{int}\left( t\right) \right] \right\rangle ,
\end{equation}
where
\begin{equation}
\left\langle ...\right\rangle =Tr\left( \rho \left( t\right) ...\right) .
\end{equation}
All operators are in the interaction representation.

The statistical operator $\rho\left( t\right) $ satisfies the
equation

\begin{equation}
i\hbar \frac{\partial \rho }{\partial t}=\left[ H_{int}\left( t\right) ,\rho
\left( t\right) \right] ,
\end{equation}
which can be solved using perturbation theory in $H_{int}$ (but
for the arbitrary $H_{1}$):
\begin{equation}
\;\rho \left( t\right) =\rho _{0}+\frac{1}{i\hbar }\int\limits_{-\infty
}^{t}dt^{\prime }\left[ H_{int}\left( t^{\prime }\right) ,\rho _{0}\right]
+\cdot \cdot \cdot
\end{equation}

We would then have

\begin{equation}
\Delta J=-\frac{1}{\hbar ^{2}V}\int\limits_{-\infty }^{t}dt^{\prime
}Tr\left( \rho _{0}\left[ \left[ H_{1},H_{int}(t)\right] ,H_{int}(t^{\prime
})\right] \right) .
\end{equation}

The decrease in total conductance $\Delta G=-G_{2},$ results the
quantum interference is defined as
\begin{equation}
G_{2}=-\frac{d\Delta J}{dV}.
\end{equation}
If the applied bias $eV$ is much smaller than the differences
between the energies $\varepsilon _{\beta }$ of modes, the Eq.
(15) describes the dependence of total conductance on the voltage
$V.$

\section{The conductance oscillations.}

\bigskip Now using the general Eqs. (14), (15), we investigate the
behaviour of $G_{2}$ for the case of $\delta -$ function
scattering potential. The Hamiltonian $H_{int}$ can be written as

\begin{equation}
\;H_{int}({\bf r}_{j})=g%
\mathrel{\mathop{\sum }\limits_{\alpha \neq \alpha ^{`}}}%
\varphi _{\alpha }^{\ast }\left( {\bf r}_{j}\right) \varphi _{\alpha
^{\prime }}\left( {\bf r}_{j}\right) c_{\alpha }^{\dagger }c_{\alpha
^{\prime }}.
\end{equation}
Here $g$ is the coupling constant of the interaction of an
electron with an impurity located in the point ${\bf r}_{j}$.

Substituting \ Eqs.(7),(16) into Eq.(14), after some simple but
cumbersome calculations we find
\begin{eqnarray}
 \Delta J=-\frac{e\pi }{2\hbar }g^{2}\sum\limits_{\alpha ,\alpha ^{\prime
}}\sum\limits_{i,j}\left( signv_{z\alpha }-signv_{z\alpha ^{\prime }}\right)\times \; \; \; \nonumber \\
 \varphi _{\alpha ^{\prime }}^{\ast }({\bf r}_{j})\varphi _{\alpha }^{\ast }(%
{\bf r}_{i})\varphi _{\alpha ^{\prime }}({\bf r}_{i})\varphi _{\alpha }({\bf %
r}_{j})(f_{\alpha ^{\prime }}-f_{\alpha })\delta (\epsilon _{\alpha ^{\prime
}}-\epsilon _{\alpha }),
\end{eqnarray}
where $f_{\alpha }=f_{F}\left( \varepsilon
+\frac{eV}{2}signv_{z\alpha }\right) $. At zero temperature
$f_{F}=\Theta \left( \varepsilon _{F}-\varepsilon \right) ,$ the
Eq.(17) can be further simplified. Using the wave functions (4),
we obtain for nonlinear part of conductance the following
equation:
\end{multicols}
\begin{eqnarray}
G_{2}& =G_{0}\frac{\pi m_{e}g^{2}}{2}\sum_{\beta ,\beta ^{\prime
},i,j}\left\{ \cos \left[ \frac{1}{\hbar }\left( p_{\beta }^{\left( +\right)
}+p_{\beta ^{\prime }}^{\left( +\right) }\right) \Delta z_{ij}\right] \frac{1%
}{p_{\beta }^{\left( +\right) }p_{\beta ^{\prime }}^{\left( +\right) }}%
\Theta \left( \varepsilon _{F}-\varepsilon _{\beta }+\frac{eV}{2}\right)
\times \right.  \\
& \Theta \left( \varepsilon _{F}-\varepsilon _{\beta ^{\prime }}+\frac{eV}{2}%
\right) +\cos \left[ \frac{1}{\hbar }\left( p_{\beta }^{\left( -\right)
}+p_{\beta ^{\prime }}^{\left( -\right) }\right) \Delta z_{ij}\right] \frac{1%
}{p_{\beta }^{\left( -\right) }p_{\beta ^{\prime }}^{\left( -\right) }}%
\Theta \left( \varepsilon _{F}-\varepsilon _{\beta }-\frac{eV}{2}\right)
\times   \nonumber \\
& \left. \Theta \left( \varepsilon _{F}-\varepsilon _{\beta n^{\prime }}-%
\frac{eV}{2}\right) \right\} \psi _{\beta ^{\prime }}^{\ast }({\bf R}%
_{j})\psi _{\beta }^{\ast }({\bf R}_{i})\psi _{\beta ^{\prime }}({\bf R}%
_{i})\psi _{\beta }({\bf R}_{j}).  \nonumber
\end{eqnarray}
\begin{multicols}{2}
Where
\begin{equation}
p_{\beta }^{\left( \pm \right) }=\sqrt{2m_{e}\left( \varepsilon _{F}\pm
\frac{eV}{2}-\varepsilon _{\beta }\right) .}
\end{equation}

The cosine terms in the Eq. (18) describe the conductance
oscillations due to the interference of electrons waves scattered
by impurities. The transverse parts $\psi _{\beta }({\bf R}_{j})$
of wave functions contain the mesoscopic effect of impurity
positions inside the constriction. The equation (18) diverges at
$p_{\beta }^{\left( \pm \right) }=0.$ Physically it means that in
the Born approximation the slowly moving electron is repeatedly
scattered on the impurity. In this case the perturbation theory
(Born approximation) is not valid any more, and we must take into
account the interference of partial waves under the electron
scattering by impurity. We assume that energy levels are not very
close to
the boundary energies $\varepsilon _{F}\pm \frac{eV}{2}$ and the quantity $%
G_{2}$ added to the ballistic conductance $G_{1}$ (8) is small.

\ For the numerical calculations we have used the free electron
model of point contact consisting of two infinite half-spaces
connected by a long ballistic cylinder of a radius $R$ and a
length $L$ (Fig.1).

In the limit $L\rightarrow\infty$ the electron wave functions
$\varphi _{\alpha^{\prime}}\left( {\bf r}\right) $ and energies
$\varepsilon _{\alpha}$ can be written as

\begin{equation}
\varphi_{\alpha^{\prime}}\left( {\bf r}\right) =\frac{1}{\sqrt{\Omega }%
J_{m+1}\left( \gamma_{mn}\right) }J_{m}\left( \gamma_{mn}\frac{\rho}{R}%
\right) \exp\left( im\varphi+\frac{i}{\hbar}p_{z}z\right) ;
\end{equation}

where

\begin{equation}
\varepsilon _{\alpha }=\varepsilon _{mn}+\frac{p_{z}^{2}}{2m_{e}};\quad
\varepsilon _{mn}=\frac{\hbar ^{2}}{2m_{e}R^{2}}\gamma _{mn}^{2}
\end{equation}
We have used cylindrical coordinates ${\bf r=}\left( \rho ,\varphi
,z\right) $ with $z$ along the axis of cylinder. Here
$a=(n,m,p_{z})$ are
the quantum numbers, $\Omega =\pi R^{2}L$ is the volume of the channel, $%
\gamma _{mn}$ are the n-th zero of the Bessel function $J_{m}.$
Since the electron energy has degeneracy for azimuthal quantum
number $m$ ( as a result of the symmetry of the model), quantum
modes with $\pm m$ have the same contribution in conductance. In
this model, Sharvin conductance has not only steps $G_{0},$ but
also steps $2G_{0}$ \cite{3Dtheor1}. In Fig.2 the dependence of
the interference pattern on the number of impurities inside a
constriction with constant radius is shown. It shows that as a
result of the interference of electron waves, which were scattered
by different impurities, the interference maxima in $G_{2}\left(
V\right) $ dependence, may both be depressed and increased. The
interference oscillation of the conductance depends strongly on
the number of opened quantum modes which follows from the
dependence of its maximum value of longitudinal electron momentum
(see, Eq. 3) on the contact size. The voltage dependence of $\
G_{2}$ for different contact sizes are shown in Fig.3. Fig. 4.
illustrates how the changing in the nonlinear dependence
$G_{2}\left( V\right) $ changes with the contact size. It
corresponds to the case of two impurities in the contact. The
position of impurities and the number of opened quantum modes are
kept constant. The difference in the interference oscillations is
a result of the changing in the relative positions of nodes and
maxima of the electron wave function from the points, in which the
impurities are situated.

\section{Conclusion}

The dependence of quantum conductance of metallic ultrasmall contacts
containing impurities on bias voltage has been theoretically studied. We
have shown that impurities situated inside the quantum microconstriction
produce a nonlinear dependence of the conductance on the applied voltage,
which is the result of the interference of electron waves reflected by
impurities. The transmission probability of the electron through
constriction depends on the relation between the electron wave length $%
\lambda $ and $\Delta z_{ij},$ the projection of distances between
impurities along the channel$.$ It is maximum when the condition
$\Delta z_{ij}=\frac{n\lambda }{2}$ ( $n$ is integer) is
satisfied. Since the electron momentum depends on the applied
bias, one can change the transmission by changing the voltage. Our
numerical calculations show that the resulting nonmonotonic
dependence of the conductance, is similar in shape to the ones
observed in experiment \cite{Rutenbeek,Agrait}. The amplitude of
interference pattern is sensitive to the transversal position of
impurities inside the constriction. If the impurity is located
near the point, where the electron wave function corresponding to
the $n$th quantum mode vanishes, then the decreasing of
transmission of that mode would be negligible.

We acknowledge fruitful discussion with M.R.H. Khajehpour and
I.K.Yanson.

\newpage
\end{multicols}
\begin{figure}
\epsfysize=1.6truein
\centerline{\epsffile{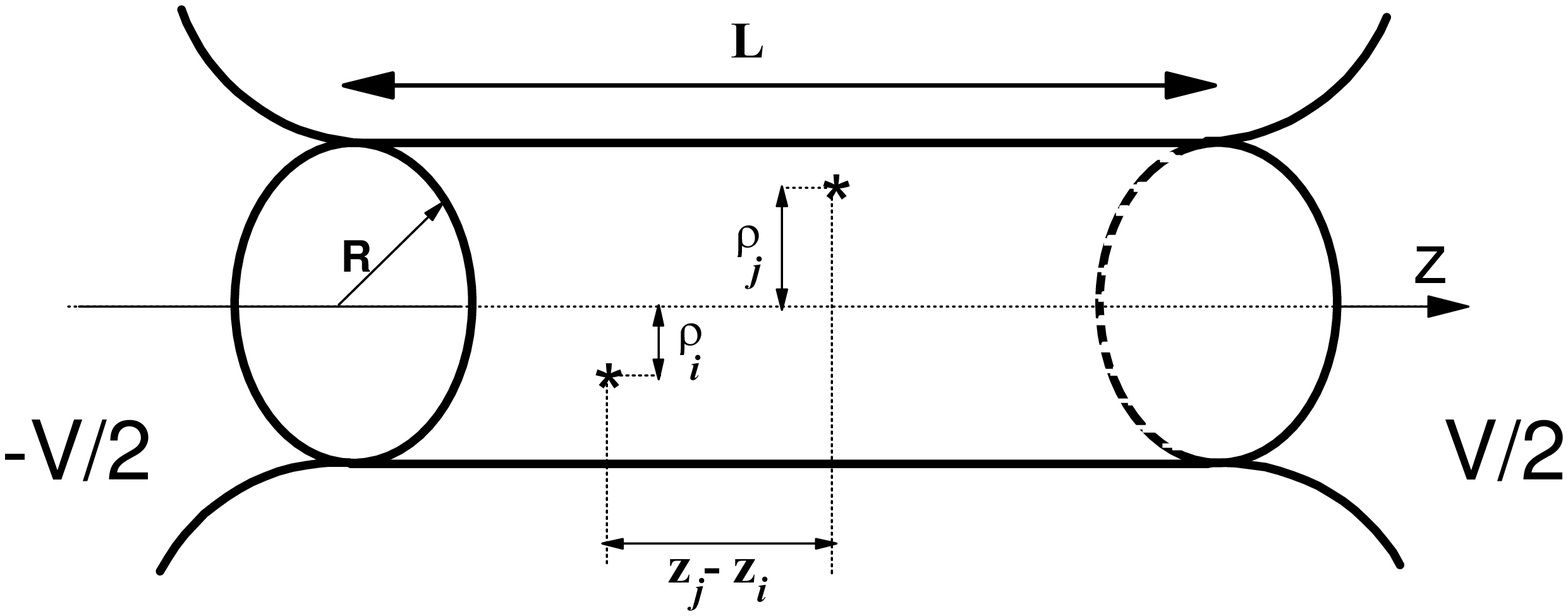}}
FIG.~1. The model of the quantum microconstriction in the form of a
long channel of the radius $R,$ which smoothly (on the Fermi length scale)
connects two massive metallic reservoirs. The impurities inside the
constriction are shown schematically.
\label{Fig1}
\end{figure}
\vspace{-2cm}

\begin{figure}
\epsfysize=4.0truein
\centerline{\epsffile{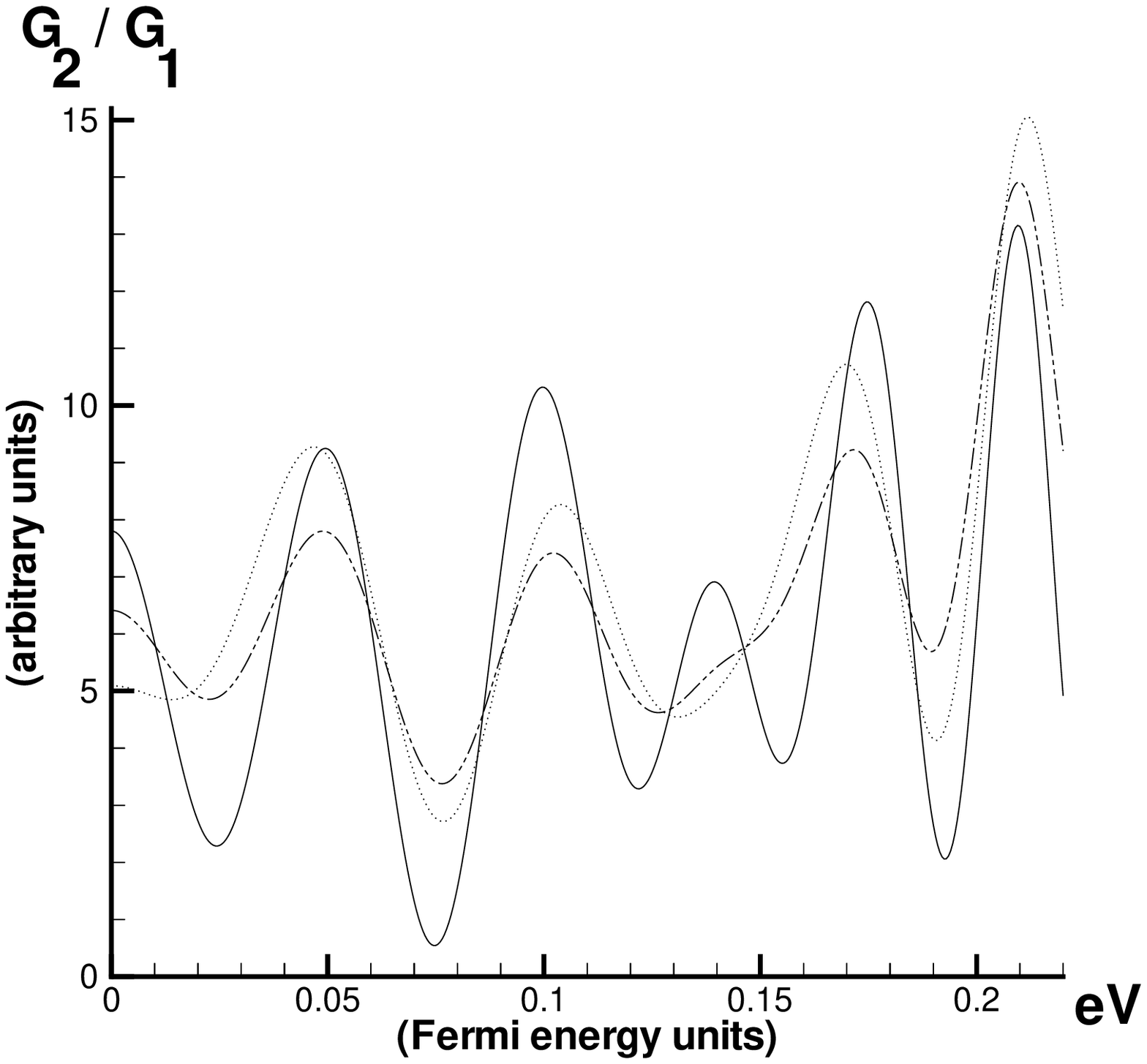}}
\vspace{1cm}
FIG.~2. The normalized conductance $G_{2}/G_{1}$ as a function of voltage in
a three modes channel ($2\pi R=4.2\lambda _{F},$) for the different number $%
k $ of impurities; $k=2$ for solid line, $k=3$ for dashed line and $k=4$ for
dotted line.
\label{Fig2}
\end{figure}
\newpage
\begin{figure}
\epsfysize=4.0truein
\centerline{\epsffile{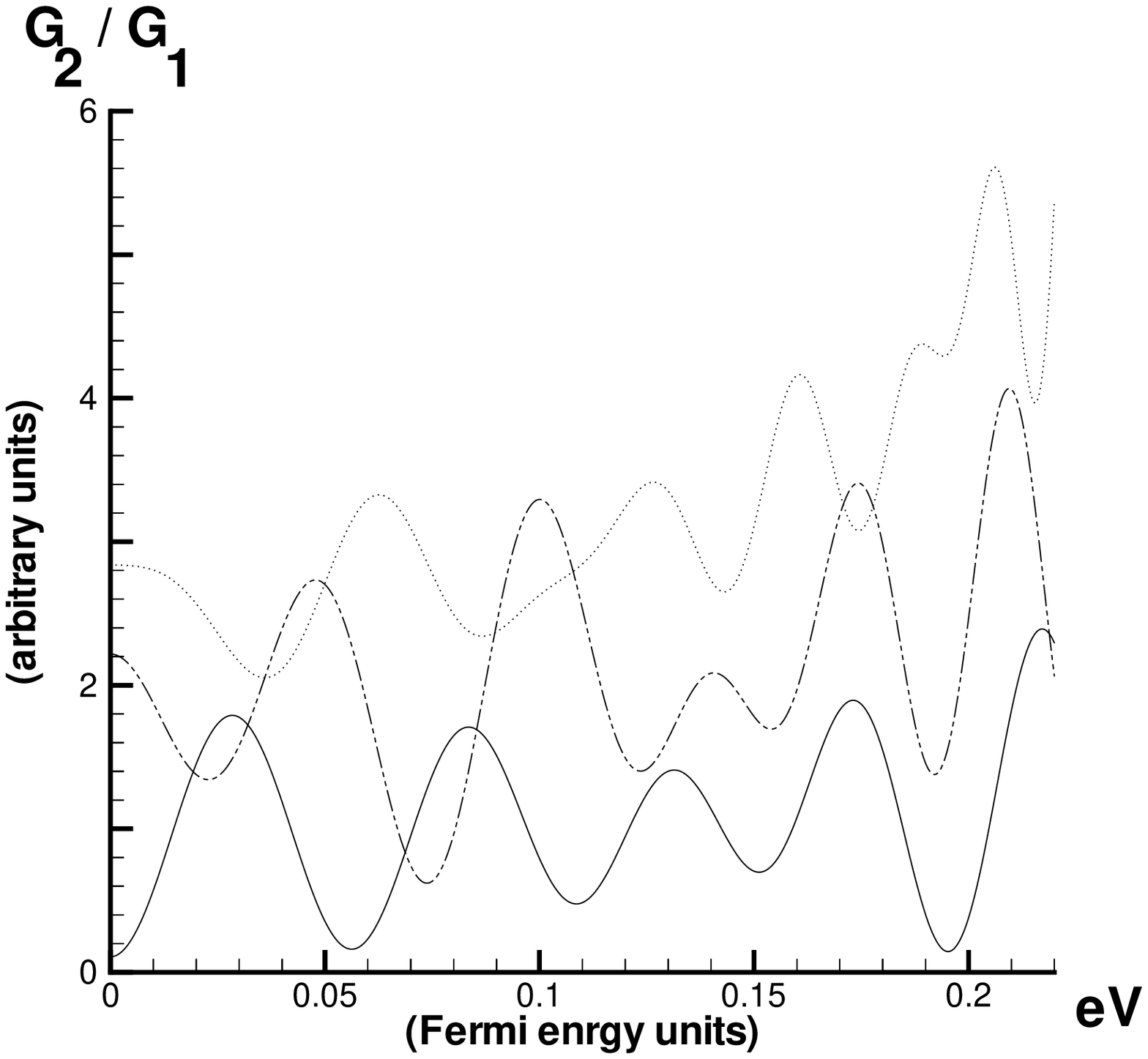}}
\vspace{1.0cm}
FIG.~3. The normalized conductance $G_{2}/G_{1}$ as a function of voltage
for a channel with two impurities at different number of opened quantum
modes; single mode $\left( 2\pi R=2.7\lambda _{F}\right) ,$ for solid line,
three modes $\left( 2\pi R=4.2\lambda _{F}\right) ,$ for dashed line and
five modes $\left( 2\pi R=5.5\lambda _{F}\right) $ for dotted line.
\label{Fig3}
\end{figure}
\begin{figure}
\epsfysize=3.0truein
\centerline{\epsffile{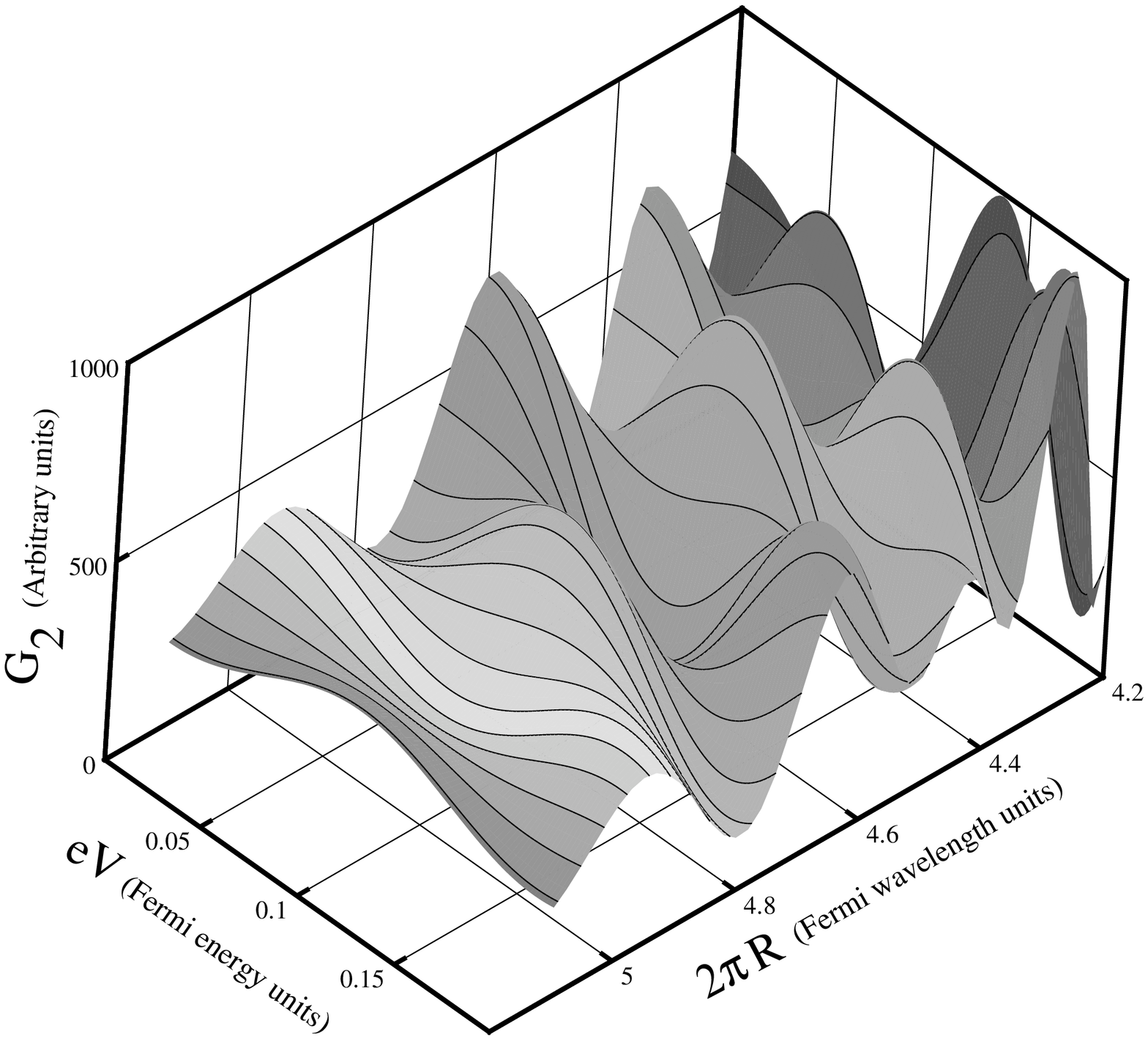}}
FIG.~4. The changing of the interference pattern in the $G_{2}\left(
V\right) $ dependence of three mode channel, which contains two impurities,
with increasing the contact diameter. The distance between impurities and
its distances from the contact axis is the same for all values of $R.$ The
radius $R$ is changed in the interval, in which the number of opened quantum
modes is constant.
\label{Fig4}
\end{figure}

\end{document}